\documentclass[PRL,aps,twocolumn,superscriptaddress,10pt,showpacs,longbibliography,noeprint]{revtex4-1}
\usepackage{graphicx}
\usepackage{color}
\usepackage{xfrac}
\usepackage{amsmath,amssymb}
\usepackage{fixmath}

\usepackage[english]{babel}
\usepackage[autostyle, english = american]{csquotes}
\MakeOuterQuote{"}
\usepackage{textgreek}
\usepackage[export]{adjustbox}
\usepackage{siunitx}
\usepackage{xcolor}
\usepackage{hyperref}
\usepackage{notes2bib}
\usepackage[version=4]{mhchem} 
\usepackage{gensymb} 
\usepackage{xfrac} 

\begin{document}


\title{Structural origins of the infamous "Low Temperature Orthorhombic" to "Low Temperature
Tetragonal" phase transition in high-T\textsubscript{c} cuprates}	

\author{Jeremiah P. Tidey} 
\affiliation{Department of Chemistry, University of Warwick, Gibbet Hill, Coventry, CV4 7AL, United Kingdom}

\author{Christopher Keegan}
\affiliation{Departments of Physics and Materials, and the Thomas Young Centre, Imperial College London, London SW7 2AZ, United Kingdom}

\author{Nicholas C. Bristowe}
\affiliation{Centre for Materials Physics, Durham University, South Road, Durham, DH1 3LE, United Kingdom}

\author{Arash A. Mostofi}
\affiliation{Departments of Physics and Materials, and the Thomas Young Centre, Imperial College London, London SW7 2AZ, United Kingdom}

\author{Zih-Mei Hong}
\affiliation{Center for Condensed Matter Sciences and Center of Atomic Initiative for New Materials, National Taiwan University, Taipei 10617, Taiwan}
\affiliation{Department of Chemistry, Fu-Jen Catholic University, New Taipei 24205, Taiwan}

\author{Bo-Hao Chen}
\affiliation{National Synchrotron Radiation Research Center, Hsinchu 30076, Taiwan}

\author{Yu-Chun Chuang}
\affiliation{National Synchrotron Radiation Research Center, Hsinchu 30076, Taiwan}

\author{Wei-Tin Chen} 
\affiliation{Center for Condensed Matter Sciences and Center of Atomic Initiative for New Materials, National Taiwan University, Taipei 10617, Taiwan}
\affiliation{Taiwan Consortium of Emergent Crystalline Materials, Ministry of Science and Technology, Taipei 10622 Taiwan}

\author{Mark S. Senn}
\email{m.senn@warwick.ac.uk}
\affiliation{Department of Chemistry, University of Warwick, Gibbet Hill, Coventry, CV4 7AL, United Kingdom}
 
\date{\today}

 \begin{abstract}
We undertake a detailed high-resolution diffraction study of a novel plain band insulator, \ce{La2MgO4}, which may be viewed as a structural surrogate system of the undoped end-member 
of the high-T\textsubscript{c} superconductors, \ce{La_{2-x-y}A^{2+}_xRE^{3+}_yCuO_{4}} \mbox{(A = \ce{Ba}, \ce{Sr}, RE
= Rare Earth)}. We find that \ce{La2MgO4} exhibits the infamous low-temperature orthorhombic (LTO) to low-temperature tetragonal (LTT) phase transition that has been linked to the suppression of 
superconductivity in a variety of underdoped cuprates, including the well known
\ce{La_{2-x}Ba_{x}CuO4} ($x=0.125$). Furthermore, we find 
that the LTO-to-LTT phase transition in \ce{La2MgO4} occurs for an octahedral tilt angle 
in the $4$\degree\ to $5$\degree\ range, similar to that which has previously been identified as a critical tipping point for superconductivity in these systems. We show that this phase transition,
occurring in a system lacking spin correlations and competing electronic states such as charge-density waves and superconductivity,
can be understood by simply navigating the density-functional theory ground-state energy landscape as a function of the order parameter amplitude.
This result calls for a careful re-investigation of the origins of the phase transitions in high-T\textsubscript{c} superconductors based on the hole-doped, $n = 1$ Ruddelsden-Popper lanthanum cuprates.
\end{abstract}

\maketitle

High-temperature superconductors based on \ce{La_{2-x}Ba_{x}CuO4} (LBCO), in which
unconventional superconductivity was first reported \cite{Bednorz1986}, continue to
receive significant attention for their intertwined electronic orders and promise of insights towards room-temperature superconductivity (SC) \cite{Berg2009, Bussmann-Holder2019, Uchida2021}. Nevertheless, a comprehensive understanding of the origin and mechanism behind such phenomena remains elusive \cite{Keimer2015}. While significant efforts have recently been made to explain the interplay of their competing electronic states \cite{Tranquada2020,Tranquada2021}, these fall short of reconciliation with the complex structural behavior concurrently observed \cite{Fausti2011a}.  

The doping-dependent temperature phase diagram of the canonical system, LBCO, exhibits
two pronounced regions of superconductivity, with peaks in the critical transition temperature
for superconductivity, T\textsubscript{c}, at \mbox{$x = 0.095$ and $0.155$}, either
side of a pronounced dip for $x\!\sim\!\frac{1}{8}$ doping \cite{Hucker2011a}. This 
suppression of 
T\textsubscript{c} is understood to be coincident \cite{Axe1994} with a structural phase transition, commencing below 80~K (T$_{\mathrm{LT}}$) from the low-temperature orthorhombic
(LTO) phase ($Bmab$) to a low-temperature tetragonal (LTT) phase
($P4_2/ncm$) \cite{Axe1989}. This phase transition has been linked to the emergence of 
a charge-density wave state \cite{Kim2008a} that is thought to compete with
three-dimensional superconductivity \cite{Tranquada1997,Moodenbaugh1998,Berg2007}. 
Since the LTT phase itself is not observed in the undoped end-member, \ce{La2CuO4}, and 
appears otherwise to be localized in the temperature, doping and composition phase 
diagram at points coincidental with suppression of T\textsubscript{c}
\cite{Suzuki1989,Cox1989,Axe1989}, it has seemed reasonable to assume that this phase arises
as a result of an interplay between electronic, spin and lattice degrees of freedom near
$x\sim\frac{1}{8}$ chemical doping \cite{Axe1994, Sakita1999}. 

\begin{figure}[!ht]
	\centering
	\includegraphics[width=\columnwidth]{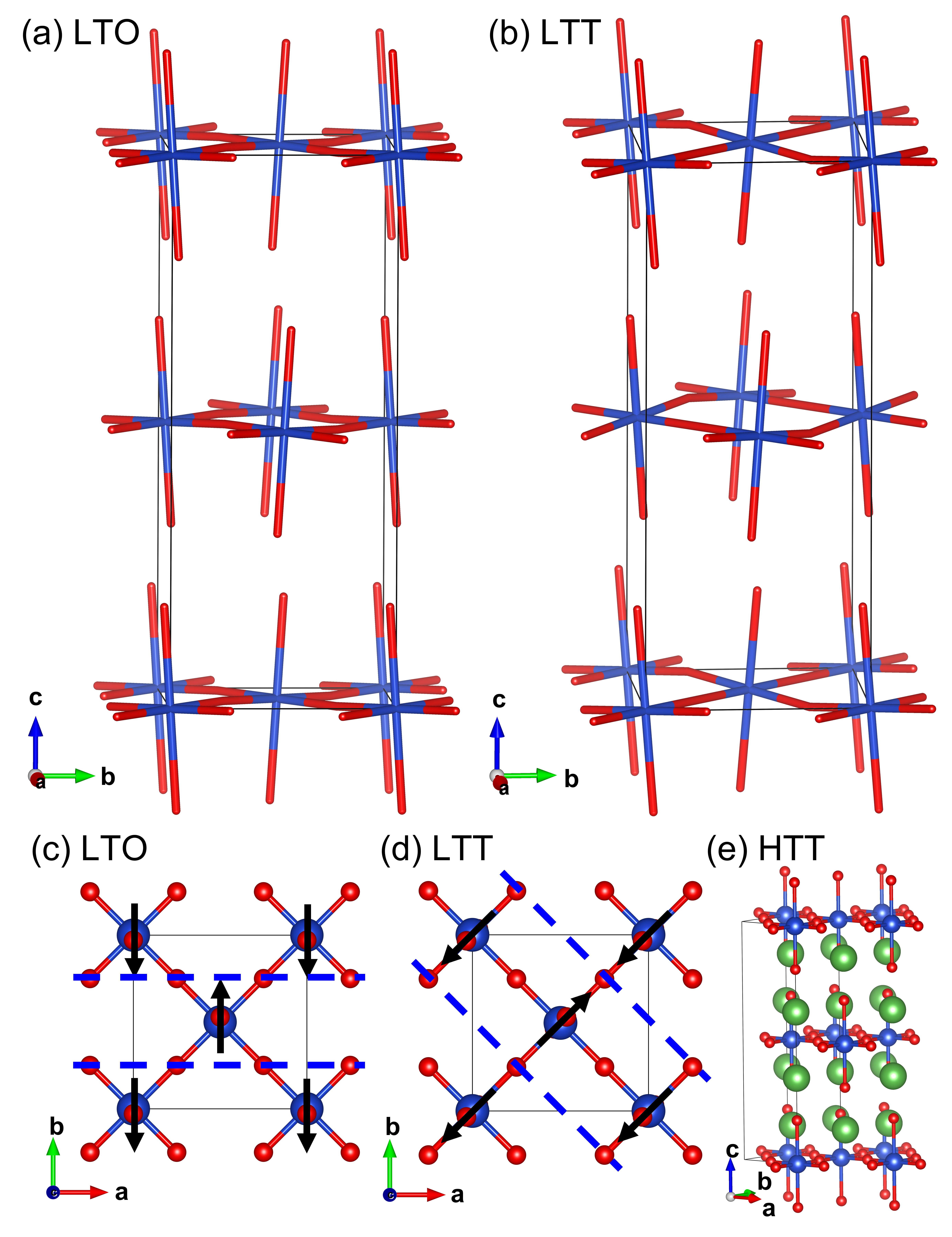}
	\caption{(a)-(d) Stick representation of the LTO and LTT structures common to the lanthanum cuprates and shared by La$_2$MgO$_4$, showing the sense of the \ce{CuO6} octahedral
	tilts and excluding A-site cations for clarity. The arrows indicate the displacement of the axial 
	oxygen atoms. (e) High-temperature tetragonal (HTT) phase with untilted octahedra with the positions
	of the La ions indicated (green spheres). Copper and oxygen atoms are depicted in
	blue and red, respectively.}
	\label{fig:LTTvsLTO}
\end{figure}

To make progress in understanding any entanglement between structural symmetries,
superconductivity and charge density waves, it is important to understand the precise
symmetries and structural mechanisms implicated in their behavior. Indeed, even in the
undoped end-member, \ce{La2CuO4}, the true structure and its physical origin remains a 
topic of discussion \cite{Billinge1994, Reehuis2006, Sapkota2021}. To gain insight
into this question, we synthesize a novel surrogate of the end-member, \ce{La2MgO4}, in
which the possibility of electronic contributions to the phase behavior is eliminated.

Polycrystalline \ce{La2MgO4} was prepared via high pressure synthesis from stoichiometric
amounts of \ce{La2O3} (Alfa Aesar, 99.999\%, dried by heating at 1273 K for 12 hours)
and \ce{MgO} (Acros, 99.99\%). The starting materials were mixed thoroughly, sealed in
a gold capsule and heated at 1273~K at
6 GPa for 30 minutes in a DIA-type cubic anvil high pressure apparatus. The sample was
quenched to room temperature after the heating program and the pressure released
slowly. \ce{La2MgO4} adopts a Ruddlesden-Popper (RP) $n = 1$ structure consisting of single
\ce{MgO6} perovskite slabs separated by \ce{LaO} rock-salt layers (Fig.~\ref{fig:LTTvsLTO}). 
As \ce{Mg^{2+}} has almost the same ionic radius as \ce{Cu^{2+}} (0.72~\AA\ vs. 0.73~\AA, respectively)
\cite{Shannon1976a}, it may be viewed as a structural surrogate to the end-member
of the RP $n = 1$ superconducting cuprate \ce{La2CuO4}, but with the key difference being that
necessarily lacks the electron-electron correlations from which competing emergent phenomena may arise.

\begin{figure}[!ht]
	\centering
	\includegraphics[width=\columnwidth]{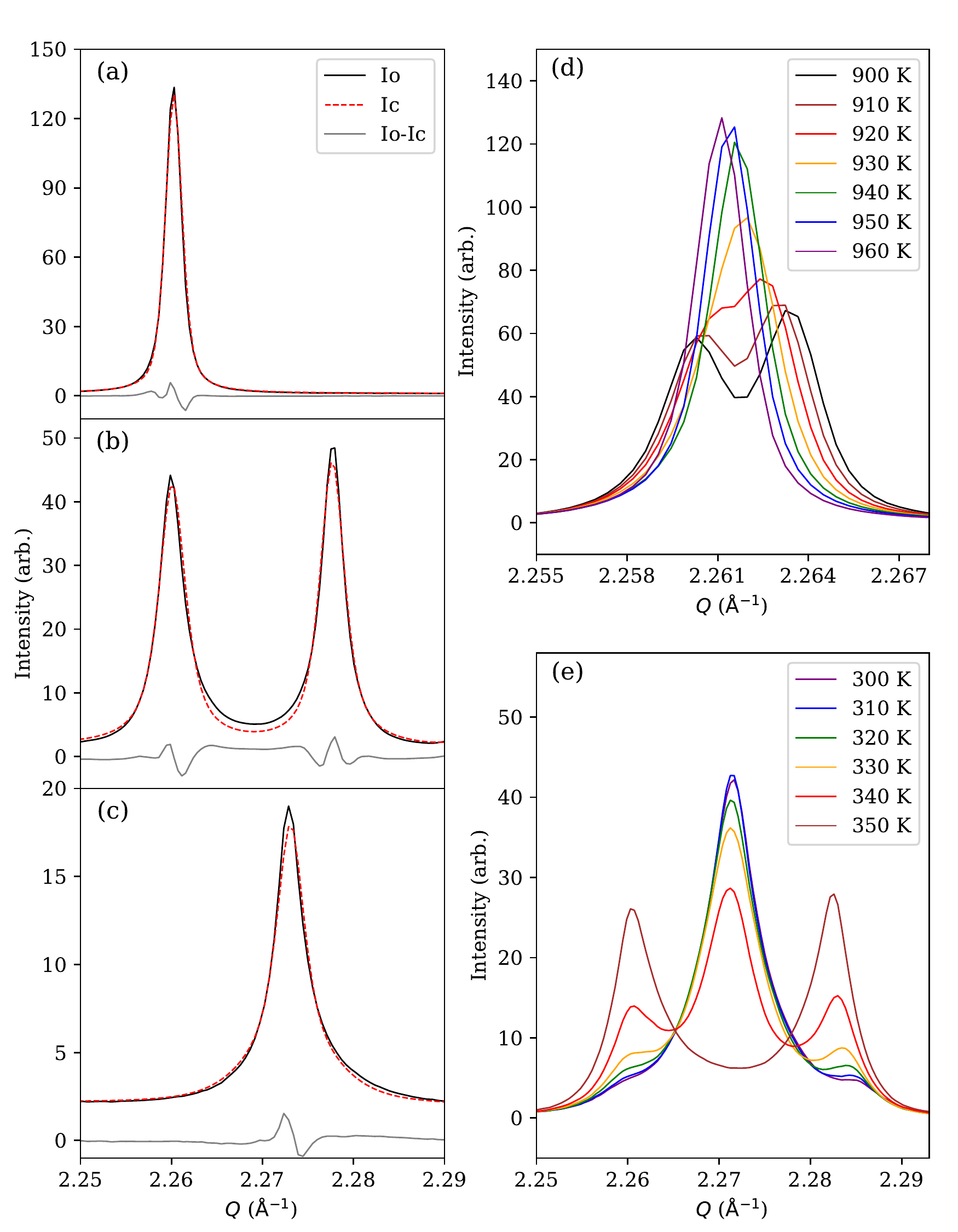}
	\caption{A plot of the Rietveld fit to the (a) $(1 1 0)_{\mathrm{HTT}}$
	diffraction profile which becomes (b) $(2 0 0)_{\mathrm{LTO}}$ / $(0 2 0)_{\mathrm{LTO}}$ and further (c)
	$(2 0 0)_{\mathrm{LTT}}$, shown at 990, 500 and 100~K, respectively. In all cases a single phase model is 
	used to fit the data. Figures (d) \& (e) show the evolution of peak profiles across the
	LTO-to-HTT and LTT-to-LTO phase transitions, evidencing second- and first-order behavior, respectively.}
	\label{fig:peaks}
\end{figure}

To ascertain the temperature-dependent structural phase diagram, synchrotron X-ray 
powder diffraction measurements were conducted on beamline 19A of the Taiwan 
Photon Source, National Synchrotron Radiation Research Center (TPS, NSRRC) with a 20~keV
and 16~keV incident beam ($\lambda = 0.61994$~\AA\ and $\lambda = 0.77495$~\AA) for the 
low- and high-temperature sweeps, respectively, using a MYTHEN detector. The \ce{La2MgO4}
sample was sealed in a 0.1~mm borosilicate capillary for low temperature and 0.1~mm
quartz capillary for high-temperature experiments, each kept spinning during data
collection for better averaging. A Cryostream 800 Plus (Oxford Cryosystems) was
utilized for data collection from 100~K to 450~K with intervals of 50~K (omitting
400~K). A hot air gas blower (FMB Oxford) was utilized for data collection from
300~K to 1000~K, at intervals of 10~K from 300~K to 370~K, 50~K from
400~K to 500~K, and 10~K from 510~K to 1000~K. All measurements were taken upon
warming. Rietveld refinement of the data was performed in Topas 6 \cite{Coelho2018}
using the symmetry-adapted displacement formalism \cite{Campbell}, as implemented in
ISODISTORT \cite{Campbell2006}. All refinements were performed in the 
(\textbf{a}+\textbf{b}, \textbf{a}-\textbf{b}, \textbf{c}) supercell with respect to the $I4/mmm$, HTT aristotype phase, and in the common subgroup $Pccn$ with appropriate
constraints imposed on the symmetry-adapted displacements and lattice parameters 
to reproduce the symmetry of the HTT ($F4/mmm$ in the aforementioned supercell), LTO or LTT
phase. For each phase, the scale factor, isotropic thermal displacement parameters
(for each atom type), Stephens strain parameters \cite{Stephens1999a}, and atomic
positions as described above were fitted, along with impurity phases as detailed in
the Supplemental Information (SI) \cite{SI}. For the temperature range 300~K - 340~K of the
high-temperature sweep, we observed clear phase coexistence of the LTO and LTT 
phases (Fig.~\ref{fig:peaks}), although a model reliable enough for structural discussion
could not be obtained until 370~K. Further details and representative fits are given in the SI.

\begin{figure}[!ht]
	\centering
	\includegraphics[width=\columnwidth]{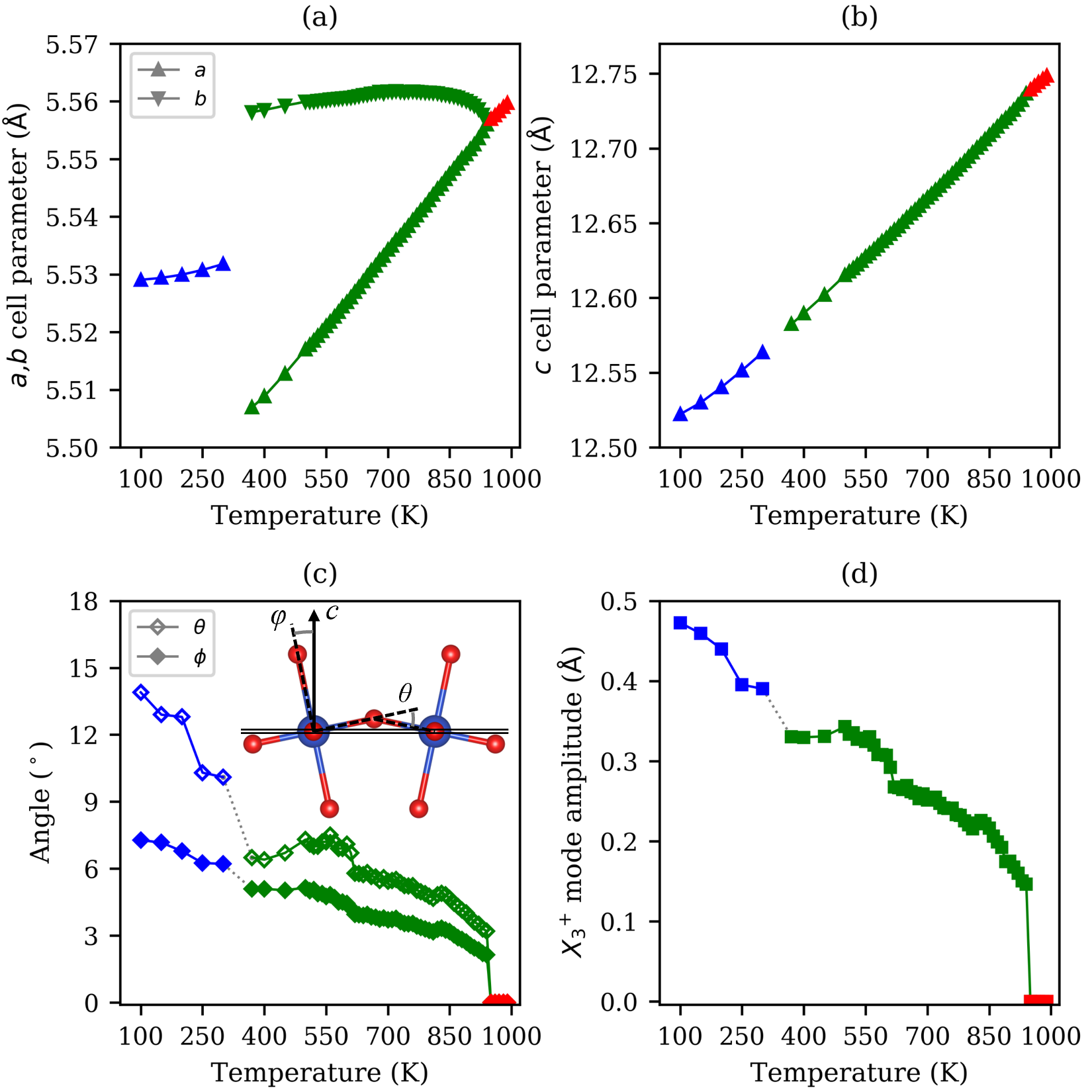}
	\caption{Parameters extracted by Rietveld refinement against variable temperature powder diffraction data collected on \ce{La2MgO4}. Colors denote LTT (blue), LTO (green) and HTT ($F4/mmm$ setting; red) single phase regimes. HTT cell parameters are reported in the supercell setting for clarity. Angles of the octahedral tilt and Mg-O-Mg buckle are determined from the refined atomic positions. The X$_3^+$ OP mode amplitude is the total mode amplitude A\textsubscript{p} value as defined by ISODISTORT \cite{Campbell2006}. A discussion of the errors in this plot can be found in the SI.}
	\label{fig:prms}
\end{figure}

From 950~K, the aristotypical $n = 1$ RP symmetry is observed with space group $F4/mmm$ in the
supercell setting, with any splitting of the $(2 0 0)$ diffraction profile clearly absent
(Fig.~\ref{fig:peaks}(a)). Below 950~K, a second-order phase transition occurs, evidenced by a
continuous evolution of the lattice parameters (Fig.~\ref{fig:prms}). This phase is
well fit by the $Bmab$ (LTO) phase (Fig.~\ref{fig:LTTvsLTO}) that is nominally the ground-state
structure of \ce{La2CuO4} \cite{Jorgensen1987}. The diffraction profile of the $(2 0 0)$
peak shows a continuous splitting into well-resolved $(2 0 0)$ and $(0 2 0)$ reflections.
This phase transition may be described as an out-of-phase rotation of the \ce{MgO6}
octahedra about the \textbf{a}-axis. Following the conventions of ISODISTORT
\cite{Campbell2006}, the order parameter (OP) associated with this tilting may be labeled
as transforming as the irreducible representation (irrep) X$_{3}^{+}$(a;0), while the tilt system of the perovskite slabs can be labeled in Glazer notation as  \textbf{a}$^-$\textbf{a}$^-$\textbf{c}$^0$ with respect to the lattice vectors of the aristotypical HTT cell.

Below 350~K, a third distinct phase is observed with tetragonal symmetry and fitting well to the LTT phase that
has been much discussed in the underdoped cuprates, most notably in LBCO ($x \approx \frac{1}{8}$). This corresponds now to
alternating tilting of the octahedra about the $[1 1 0]$ and $[1 \bar{1} 0]$ axes in the
perovskite layers, centered at $z$ and $z$ + $\frac{1}{2}$, and may be described by
an OP transforming as the irrep X$_3^+$(a;a) (\textbf{a}$^-$\textbf{a}$^0$\textbf{c}$^0$).Upon warming
through this transition, a clear coexistence between the LTT and LTO phases
is observed (Fig.~\ref{fig:peaks}(e)). Along with a discontinuous jump in the lattice
parameters and refined magnitude of the OP (Fig.~\ref{fig:prms}), this is indicative of first-order behavior. 
Since this phase transition has invariably been found to be concomitant with the suppression of superconductivity
and the formation of CDWs, the novel observation of the LTT phase in a system that is a plain band insulator, i.e., lacking the possibility of strong electronic
correlations, allows opportunity for clear insight into the structural contributions to the mechanism of the LTO-to-LTT phase transition,
isolated from the influences of competing electronic phenomena.

To obtain further insight into the origin of the LTO-to-LTT phase transition in \ce{La2MgO4}, we
perform single-point energy calculations using density-functional theory (DFT)
to map out an energy landscape spanning the X$_3^+$(a;0) and X$_3^+$(a;a) OPs. The symmetry in all cases can be described within the $Pccn$ space group,
which is a common subgroup of both the LTT and LTO phases and encompasses intermediate tilt propagation
vectors that describe the so-called low-temperature less orthorhombic (LTLO) structure.
Figure 4(a) shows the energy landscape associated to an interpolation of the strain, X$_3^+$ OP magnitude and direction, and the $\mathrm{\Gamma}_1^+$ OP magnitude between the relaxed HTT, LTO and LTT phases (see SI).
These single-point energy calculations reveal no metastable, local minima for any intermediate $Pccn$ phases between the LTO and LTT structures.
This is further supported by full structural relaxations of intermediate $Pccn$ structures which invariably relax to the LTT structure.
Since the $Pccn$ phase has lower symmetry than the LTO and LTT phases, we would equally expect its vibrational entropy to be lower.
The lower entropy of this phase also makes a stable $Pccn$ phase unexpected with increasing temperature.
The LTO-to-LTT phase transition can then not occur via a continuous rotation of the OP that passes through the $Pccn$ phase, meaning it must be first-order.
This is consistent with the first-order nature of the phase transition evidenced experimentally (i.e., the pronounced phase coexistence; Fig. 2(e)).

\begin{figure}[!ht]
	\centering
	\includegraphics[width=\columnwidth]{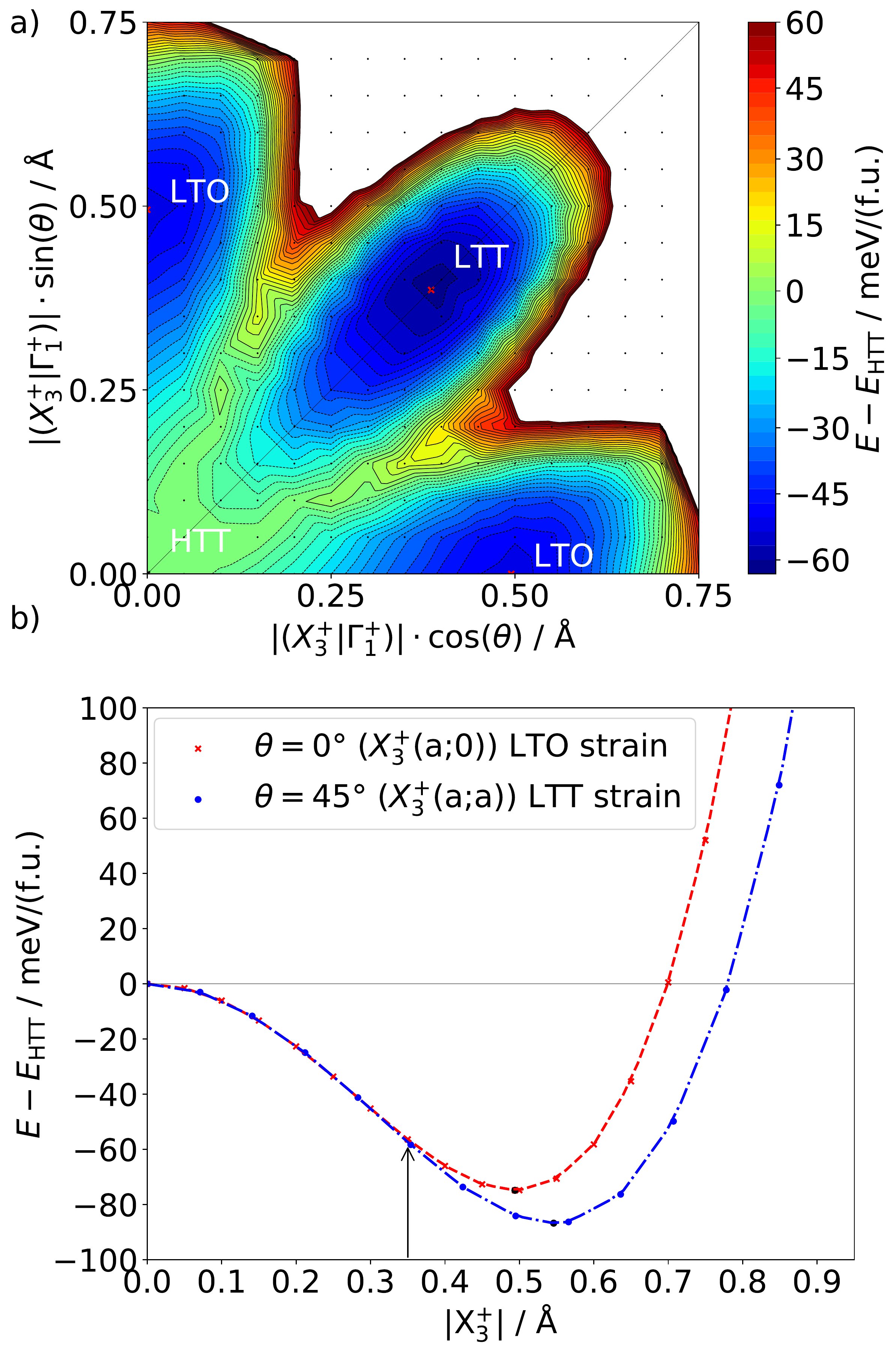}
  \caption{(a) The energy landscape for a given direction and magnitude of the X$_3^+$(a;b)
  OP with simultaneous interpolation of the $\mathrm{\Gamma}_1^+$ OP and strain field 
  between the HTT, relaxed LTO and relaxed LTT phases, calculated relative to the
  non-standard, $F4/mmm$, HTT phase (see SI for further  details).
  (b) Plot of the the energy landscape along the X$_3^+$(a;0) and X$_3^+$(a;a) OP directions at the calculated relaxed values of strain and $\mathrm{\Gamma}_1^+$ OP for the LTO and LTT phases, respectively, relative to the HTT aristotype (see SI).
  All mode amplitudes are given in A\textsubscript{p} values as defined by ISODISTORT \cite{Campbell2006}. The black data points denote the positions of the lowest-energy structures along each OP direction.}
	\label{fig:energylandscape}
\end{figure}

While our DFT calculations do not explicitly capture the effect of temperature,
we may still gain valuable insights
into the sequence of phase transitions in \ce{La2MgO4}.
Figure~\ref{fig:energylandscape}(b) shows the energy landscape with respect to the magnitude of the X$_3^+$ OP along the (a;0) and (a;a) OP directions for strain and $\mathrm{\Gamma}_1^+$ magnitudes fixed at the calculated relaxed structures for LTO and LTT, respectively.
This reveals that there is a
clear departure of the LTO and LTT phases from the harmonic approximation (which requires that all OP
directions be iso-energetic) at an amplitude of around $\mathrm{A_p} = 0.35$~\AA. Such an amplitude lies
within the discontinuous region between the LTO and LTT phases, at around 300~K to 350~K in our diffraction study
(Fig.~\ref{fig:prms}(d)).

The origin of the experimentally observed LTT-to-LTO phase transition in \ce{La2MgO4} may then be understood within the quasi-harmonic approximation
in which the thermal lattice expansion renormalizes the OP amplitude,
as we have found in our previous work on RP $n = 1$ systems in which we included the effects of thermal
expansion~\cite{Ablitt2017}. The increase in amplitude of
the OP upon cooling leads to a scenario where the LTT structure becomes more favorable from an enthalpic perspective. Any
entropic contributions to the Gibbs free energy that favor LTO at higher temperatures are evidently then
quickly overwhelmed. The observation of such a cross-over between LTO and LTT superstructures
in our surrogate system at a tilt angle between 4\degree\ and 5\degree\ may indicate
why a similar value has been identified as a critical octahedral tilt angle
for superconductivity in the family of compounds, \ce{La_{2-x-y}A^{2+}_xRE^{3+}_yCuO_{4}} \mbox{(A = \ce{Ba}, \ce{Sr}, RE
= Rare Earth)} \cite{Hucker2012}, where superconductivity is presumed to be the preserve of the LTO phase.

Finally, we turn our attention to \ce{La2CuO4} where we employ DFT with an on-site
Hubbard $U$ on the Cu-site of 9~eV (SI), an approach which has been demonstrated to be 
effective in approximating the electronic ground state of \ce{La2CuO4}
\cite{cote2011, Sterling2021}. While our
DFT+$U$ calculations for \ce{La2MgO4} find the LTT phase to be lower in energy than the LTO phase by
$\sim\!11.2$~meV/(formula unit), for \ce{La2CuO4}
this energy difference is of the order of the typical accuracy of DFT. Prior calculations 
on \ce{La2CuO4} using the SCAN exchange-correlation functional also found the two phases to be very
close in energy \cite{Furness2018}. These results
are consistent with the experimental observations that the LTO-to-LTT phase
transition is not observed upon cooling for \ce{La2CuO4} \cite{Jorgensen1987,
Billinge1994,Reehuis2006,Sapkota2021} but is evidenced here as occurring at significantly elevated
temperatures in \ce{La2MgO4}. The small energy scale associated with variation in the ionic
radii and changes in spin and electron correlations that occur as the La$_2$CuO$_4$ system is
doped towards $x = \frac{1}{8}$ may be sufficient to affect small changes in
the amplitude of the X$_3^+$ OP, in turn triggering or indeed inhibiting the transition from LTO to LTT in these systems. This
resulting change in structural symmetry is what has often been correlated with the suppression of superconductivity in these systems.
 
In conclusion, we have shown that the infamous low-temperature tetragonal (LTT) to
low-temperature orthorhombic (LTO) phase transition, observed in many
underdoped lanthanum cuprates, may be reproduced in the surrogate plain band insulator,
\ce{La2MgO4}, which is presumably devoid of electron-electron correlation inherent to the competing superconducting and CDW states. The transition is shown to be triggered simply by the change in order
parameter magnitude associated with the tilt of the \ce{BO6} octahedra (B = \ce{Mg}
or \ce{Cu}). In the doped \ce{La2CuO4}, small changes in magnetic and electronic
correlations (experimentally or {\it in silico}) will couple to the order parameter
magnitude, consequently selecting either LTO or LTT as the ground state, which
presumably themselves favor superconductivity or CDW formation, respectively. It is noteworthy
that the tilt angle at the relevant OP magnitude lies close to the critical value of tilting that has been associated to the suppression of superconductivity in the doped lanthanum cuprates.

\section*{Acknowledgements}

This work was partly supported by Ministry of Science and Technology (Taiwan) (MOST-108-2112-M-002-025-MY3) and Academia Sinica project number AS-iMATE-109-13. The synchrotron experiment was supported by National Synchrotron Radiation Research Center (NSRRC, Taiwan) with proposal no. 2021-1-024. J.P.T. was funded by EPSRC grant EP/S027106/1. M.S.S. acknowledges the Royal Society for a fellowship (UF160265). C.K. was supported through a studentship in the Centre for Doctoral Training on Theory and Simulation of Materials at Imperial College London funded by the EPSRC (EP/S515085/1 and EP/L015579/1). We acknowledge the Thomas Young Centre under grant number TYC-101.

\bibliography{references.bib}	

\begin{thebibliography}{33}%
\makeatletter
\providecommand \@ifxundefined [1]{%
 \@ifx{#1\undefined}
}%
\providecommand \@ifnum [1]{%
 \ifnum #1\expandafter \@firstoftwo
 \else \expandafter \@secondoftwo
 \fi
}%
\providecommand \@ifx [1]{%
 \ifx #1\expandafter \@firstoftwo
 \else \expandafter \@secondoftwo
 \fi
}%
\providecommand \natexlab [1]{#1}%
\providecommand \enquote  [1]{``#1''}%
\providecommand \bibnamefont  [1]{#1}%
\providecommand \bibfnamefont [1]{#1}%
\providecommand \citenamefont [1]{#1}%
\providecommand \href@noop [0]{\@secondoftwo}%
\providecommand \href [0]{\begingroup \@sanitize@url \@href}%
\providecommand \@href[1]{\@@startlink{#1}\@@href}%
\providecommand \@@href[1]{\endgroup#1\@@endlink}%
\providecommand \@sanitize@url [0]{\catcode `\\12\catcode `\$12\catcode
  `\&12\catcode `\#12\catcode `\^12\catcode `\_12\catcode `\%12\relax}%
\providecommand \@@startlink[1]{}%
\providecommand \@@endlink[0]{}%
\providecommand \url  [0]{\begingroup\@sanitize@url \@url }%
\providecommand \@url [1]{\endgroup\@href {#1}{\urlprefix }}%
\providecommand \urlprefix  [0]{URL }%
\providecommand \Eprint [0]{\href }%
\providecommand \doibase [0]{http://dx.doi.org/}%
\providecommand \selectlanguage [0]{\@gobble}%
\providecommand \bibinfo  [0]{\@secondoftwo}%
\providecommand \bibfield  [0]{\@secondoftwo}%
\providecommand \translation [1]{[#1]}%
\providecommand \BibitemOpen [0]{}%
\providecommand \bibitemStop [0]{}%
\providecommand \bibitemNoStop [0]{.\EOS\space}%
\providecommand \EOS [0]{\spacefactor3000\relax}%
\providecommand \BibitemShut  [1]{\csname bibitem#1\endcsname}%
\let\auto@bib@innerbib\@empty
\bibitem [{\citenamefont {Bednorz}\ and\ \citenamefont
  {M{\"{u}}ller}(1986)}]{Bednorz1986}%
  \BibitemOpen
  \bibfield  {author} {\bibinfo {author} {\bibfnamefont {J.~G.}\ \bibnamefont
  {Bednorz}}\ and\ \bibinfo {author} {\bibfnamefont {K.~A.}\ \bibnamefont
  {M{\"{u}}ller}},\ }\bibfield  {title} {\enquote {\bibinfo {title} {{Possible
  High Tc Superconductivity in Ba-La-Cu-O System}},}\ }\href@noop {} {\bibfield
   {journal} {\bibinfo  {journal} {Z. Phys. B - Condensed Matter}\ }\textbf
  {\bibinfo {volume} {64}},\ \bibinfo {pages} {189--193} (\bibinfo {year}
  {1986})}\BibitemShut {NoStop}%
\bibitem [{\citenamefont {Berg}\ \emph {et~al.}(2009)\citenamefont {Berg},
  \citenamefont {Fradkin}, \citenamefont {Kivelson},\ and\ \citenamefont
  {Tranquada}}]{Berg2009}%
  \BibitemOpen
  \bibfield  {author} {\bibinfo {author} {\bibfnamefont {Erez}\ \bibnamefont
  {Berg}}, \bibinfo {author} {\bibfnamefont {Eduardo}\ \bibnamefont {Fradkin}},
  \bibinfo {author} {\bibfnamefont {Steven~A.}\ \bibnamefont {Kivelson}}, \
  and\ \bibinfo {author} {\bibfnamefont {John~M.}\ \bibnamefont {Tranquada}},\
  }\bibfield  {title} {\enquote {\bibinfo {title} {{Striped superconductors:
  How spin, charge and superconducting orders intertwine in the cuprates}},}\
  }\href {\doibase 10.1088/1367-2630/11/11/115004} {\bibfield  {journal}
  {\bibinfo  {journal} {New Journal of Physics}\ }\textbf {\bibinfo {volume}
  {11}} (\bibinfo {year} {2009}),\ 10.1088/1367-2630/11/11/115004}\BibitemShut
  {NoStop}%
\bibitem [{\citenamefont {Bussmann-Holder}\ and\ \citenamefont
  {Keller}(2019)}]{Bussmann-Holder2019}%
  \BibitemOpen
  \bibfield  {author} {\bibinfo {author} {\bibfnamefont {Annette}\ \bibnamefont
  {Bussmann-Holder}}\ and\ \bibinfo {author} {\bibfnamefont {Hugo}\
  \bibnamefont {Keller}},\ }\bibfield  {title} {\enquote {\bibinfo {title}
  {{High-temperature superconductors: Underlying physics and applications}},}\
  }\href {\doibase 10.1515/znb-2019-0103} {\bibfield  {journal} {\bibinfo
  {journal} {Zeitschrift fur Naturforschung - Section B Journal of Chemical
  Sciences}\ }\textbf {\bibinfo {volume} {75}},\ \bibinfo {pages} {3--14}
  (\bibinfo {year} {2019})}\BibitemShut {NoStop}%
\bibitem [{\citenamefont {Uchida}(2021)}]{Uchida2021}%
  \BibitemOpen
  \bibfield  {author} {\bibinfo {author} {\bibfnamefont {Shin~Ichi}\
  \bibnamefont {Uchida}},\ }\bibfield  {title} {\enquote {\bibinfo {title}
  {{Ubiquitous charge order correlations in high-temperature superconducting
  cuprates}},}\ }\href {\doibase 10.7566/JPSJ.90.111001} {\bibfield  {journal}
  {\bibinfo  {journal} {Journal of the Physical Society of Japan}\ }\textbf
  {\bibinfo {volume} {90}},\ \bibinfo {pages} {1--15} (\bibinfo {year}
  {2021})}\BibitemShut {NoStop}%
\bibitem [{\citenamefont {Keimer}\ \emph {et~al.}(2015)\citenamefont {Keimer},
  \citenamefont {Kivelson}, \citenamefont {Norman}, \citenamefont {Uchida},\
  and\ \citenamefont {Zaanen}}]{Keimer2015}%
  \BibitemOpen
  \bibfield  {author} {\bibinfo {author} {\bibfnamefont {B.}~\bibnamefont
  {Keimer}}, \bibinfo {author} {\bibfnamefont {S.~A.}\ \bibnamefont
  {Kivelson}}, \bibinfo {author} {\bibfnamefont {M.~R.}\ \bibnamefont
  {Norman}}, \bibinfo {author} {\bibfnamefont {S.}~\bibnamefont {Uchida}}, \
  and\ \bibinfo {author} {\bibfnamefont {J.}~\bibnamefont {Zaanen}},\
  }\bibfield  {title} {\enquote {\bibinfo {title} {{From quantum matter to
  high-temperature superconductivity in copper oxides}},}\ }\href {\doibase
  10.1038/nature14165} {\bibfield  {journal} {\bibinfo  {journal} {Nature}\
  }\textbf {\bibinfo {volume} {518}},\ \bibinfo {pages} {179--186} (\bibinfo
  {year} {2015})}\BibitemShut {NoStop}%
\bibitem [{\citenamefont {Tranquada}(2020)}]{Tranquada2020}%
  \BibitemOpen
  \bibfield  {author} {\bibinfo {author} {\bibfnamefont {J.~M.}\ \bibnamefont
  {Tranquada}},\ }\bibfield  {title} {\enquote {\bibinfo {title} {{Cuprate
  superconductors as viewed through a striped lens}},}\ }\href {\doibase
  10.1080/00018732.2021.1935698} {\bibfield  {journal} {\bibinfo  {journal}
  {Advances in Physics}\ }\textbf {\bibinfo {volume} {69}},\ \bibinfo {pages}
  {437--509} (\bibinfo {year} {2020})}\BibitemShut {NoStop}%
\bibitem [{\citenamefont {Tranquada}\ \emph {et~al.}(2021)\citenamefont
  {Tranquada}, \citenamefont {Dean},\ and\ \citenamefont {Li}}]{Tranquada2021}%
  \BibitemOpen
  \bibfield  {author} {\bibinfo {author} {\bibfnamefont {John~M.}\ \bibnamefont
  {Tranquada}}, \bibinfo {author} {\bibfnamefont {Mark~P.M.}\ \bibnamefont
  {Dean}}, \ and\ \bibinfo {author} {\bibfnamefont {Qiang}\ \bibnamefont
  {Li}},\ }\bibfield  {title} {\enquote {\bibinfo {title} {{Superconductivity
  from charge order in cuprates}},}\ }\href {\doibase 10.7566/JPSJ.90.111002}
  {\bibfield  {journal} {\bibinfo  {journal} {Journal of the Physical Society
  of Japan}\ }\textbf {\bibinfo {volume} {90}},\ \bibinfo {pages} {1--11}
  (\bibinfo {year} {2021})}\BibitemShut {NoStop}%
\bibitem [{\citenamefont {Fausti}\ \emph {et~al.}(2011)\citenamefont {Fausti},
  \citenamefont {Tobey}, \citenamefont {Dean}, \citenamefont {Kaiser},
  \citenamefont {Dienst}, \citenamefont {Hoffmann}, \citenamefont {Pyon},
  \citenamefont {Takayama}, \citenamefont {Takagi},\ and\ \citenamefont
  {Cavalleri}}]{Fausti2011a}%
  \BibitemOpen
  \bibfield  {author} {\bibinfo {author} {\bibfnamefont {D.}~\bibnamefont
  {Fausti}}, \bibinfo {author} {\bibfnamefont {R.~I.}\ \bibnamefont {Tobey}},
  \bibinfo {author} {\bibfnamefont {N.}~\bibnamefont {Dean}}, \bibinfo {author}
  {\bibfnamefont {S.}~\bibnamefont {Kaiser}}, \bibinfo {author} {\bibfnamefont
  {A.}~\bibnamefont {Dienst}}, \bibinfo {author} {\bibfnamefont {M.~C.}\
  \bibnamefont {Hoffmann}}, \bibinfo {author} {\bibfnamefont {S.}~\bibnamefont
  {Pyon}}, \bibinfo {author} {\bibfnamefont {T.}~\bibnamefont {Takayama}},
  \bibinfo {author} {\bibfnamefont {H.}~\bibnamefont {Takagi}}, \ and\ \bibinfo
  {author} {\bibfnamefont {A.}~\bibnamefont {Cavalleri}},\ }\bibfield  {title}
  {\enquote {\bibinfo {title} {{Light-induced superconductivity in a
  stripe-ordered cuprate}},}\ }\href {\doibase 10.1126/science.1197294}
  {\bibfield  {journal} {\bibinfo  {journal} {Science}\ }\textbf {\bibinfo
  {volume} {331}},\ \bibinfo {pages} {189--191} (\bibinfo {year}
  {2011})}\BibitemShut {NoStop}%
\bibitem [{\citenamefont {H{\"{u}}cker}\ \emph {et~al.}(2011)\citenamefont
  {H{\"{u}}cker}, \citenamefont {{V. Zimmermann}}, \citenamefont {Gu},
  \citenamefont {Xu}, \citenamefont {Wen}, \citenamefont {Xu}, \citenamefont
  {Kang}, \citenamefont {Zheludev},\ and\ \citenamefont
  {Tranquada}}]{Hucker2011a}%
  \BibitemOpen
  \bibfield  {author} {\bibinfo {author} {\bibfnamefont {M.}~\bibnamefont
  {H{\"{u}}cker}}, \bibinfo {author} {\bibfnamefont {M.}~\bibnamefont {{V.
  Zimmermann}}}, \bibinfo {author} {\bibfnamefont {G.~D.}\ \bibnamefont {Gu}},
  \bibinfo {author} {\bibfnamefont {Z.~J.}\ \bibnamefont {Xu}}, \bibinfo
  {author} {\bibfnamefont {J.~S.}\ \bibnamefont {Wen}}, \bibinfo {author}
  {\bibfnamefont {Guangyong}\ \bibnamefont {Xu}}, \bibinfo {author}
  {\bibfnamefont {H.~J.}\ \bibnamefont {Kang}}, \bibinfo {author}
  {\bibfnamefont {A.}~\bibnamefont {Zheludev}}, \ and\ \bibinfo {author}
  {\bibfnamefont {J.~M.}\ \bibnamefont {Tranquada}},\ }\bibfield  {title}
  {\enquote {\bibinfo {title} {{Stripe order in superconducting La2-xBaxCuO 4
  (0.095<x<0.155)}},}\ }\href {\doibase 10.1103/PhysRevB.83.104506} {\bibfield
  {journal} {\bibinfo  {journal} {Physical Review B - Condensed Matter and
  Materials Physics}\ }\textbf {\bibinfo {volume} {83}},\ \bibinfo {pages}
  {41--50} (\bibinfo {year} {2011})}\BibitemShut {NoStop}%
\bibitem [{\citenamefont {Axe}\ and\ \citenamefont {Crawford}(1994)}]{Axe1994}%
  \BibitemOpen
  \bibfield  {author} {\bibinfo {author} {\bibfnamefont {J.~D.}\ \bibnamefont
  {Axe}}\ and\ \bibinfo {author} {\bibfnamefont {M.~K.}\ \bibnamefont
  {Crawford}},\ }\bibfield  {title} {\enquote {\bibinfo {title} {{Structural
  Instabilities in Lanthanum Cuprate}},}\ }\href@noop {} {\bibfield  {journal}
  {\bibinfo  {journal} {Journal of Low Temperature Physics}\ }\textbf {\bibinfo
  {volume} {95}},\ \bibinfo {pages} {271--284} (\bibinfo {year}
  {1994})}\BibitemShut {NoStop}%
\bibitem [{\citenamefont {Axe}\ \emph {et~al.}(1989)\citenamefont {Axe},
  \citenamefont {Moudden}, \citenamefont {Hohlwein}, \citenamefont {Cox},
  \citenamefont {Mohanty}, \citenamefont {Moodenbaugh},\ and\ \citenamefont
  {Xu}}]{Axe1989}%
  \BibitemOpen
  \bibfield  {author} {\bibinfo {author} {\bibfnamefont {J.~D.}\ \bibnamefont
  {Axe}}, \bibinfo {author} {\bibfnamefont {A.~H.}\ \bibnamefont {Moudden}},
  \bibinfo {author} {\bibfnamefont {D.}~\bibnamefont {Hohlwein}}, \bibinfo
  {author} {\bibfnamefont {D.~E.}\ \bibnamefont {Cox}}, \bibinfo {author}
  {\bibfnamefont {K.~M.}\ \bibnamefont {Mohanty}}, \bibinfo {author}
  {\bibfnamefont {A.~R.}\ \bibnamefont {Moodenbaugh}}, \ and\ \bibinfo {author}
  {\bibfnamefont {Youwen}\ \bibnamefont {Xu}},\ }\bibfield  {title} {\enquote
  {\bibinfo {title} {{Structural phase transformations and superconductivity in
  La$_{2-x}$Ba$_x$CuO$_4$}},}\ }\href {\doibase 10.1103/PhysRevLett.62.2751}
  {\bibfield  {journal} {\bibinfo  {journal} {Physical Review Letters}\
  }\textbf {\bibinfo {volume} {62}},\ \bibinfo {pages} {2751--2754} (\bibinfo
  {year} {1989})}\BibitemShut {NoStop}%
\bibitem [{\citenamefont {Kim}\ \emph {et~al.}(2008)\citenamefont {Kim},
  \citenamefont {Gu}, \citenamefont {Gog},\ and\ \citenamefont
  {Casa}}]{Kim2008a}%
  \BibitemOpen
  \bibfield  {author} {\bibinfo {author} {\bibfnamefont {Young-June}\
  \bibnamefont {Kim}}, \bibinfo {author} {\bibfnamefont {G.~D.}\ \bibnamefont
  {Gu}}, \bibinfo {author} {\bibfnamefont {T.}~\bibnamefont {Gog}}, \ and\
  \bibinfo {author} {\bibfnamefont {D.}~\bibnamefont {Casa}},\ }\bibfield
  {title} {\enquote {\bibinfo {title} {{X-ray scattering study of charge
  density waves in La$_{2 -- x}$Ba$_x$CuO$_4$}},}\ }\href {\doibase
  10.1103/PhysRevB.77.064520} {\bibfield  {journal} {\bibinfo  {journal}
  {Physical Review B}\ }\textbf {\bibinfo {volume} {77}},\ \bibinfo {pages}
  {064520} (\bibinfo {year} {2008})}\BibitemShut {NoStop}%
\bibitem [{\citenamefont {Tranquada}\ \emph {et~al.}(1997)\citenamefont
  {Tranquada}, \citenamefont {Axe}, \citenamefont {Ichikawa}, \citenamefont
  {Moodenbaugh}, \citenamefont {Nakamura},\ and\ \citenamefont
  {Uchida}}]{Tranquada1997}%
  \BibitemOpen
  \bibfield  {author} {\bibinfo {author} {\bibfnamefont {J.~M.}\ \bibnamefont
  {Tranquada}}, \bibinfo {author} {\bibfnamefont {J.~D.}\ \bibnamefont {Axe}},
  \bibinfo {author} {\bibfnamefont {N.}~\bibnamefont {Ichikawa}}, \bibinfo
  {author} {\bibfnamefont {A.~R.}\ \bibnamefont {Moodenbaugh}}, \bibinfo
  {author} {\bibfnamefont {Y.}~\bibnamefont {Nakamura}}, \ and\ \bibinfo
  {author} {\bibfnamefont {S.}~\bibnamefont {Uchida}},\ }\bibfield  {title}
  {\enquote {\bibinfo {title} {{Coexistence of, and competition between,
  superconductivity and charge-stripe order in
  La$_{1.6--x}$Nd$_{0.4}$Sr$_x$CuO$_4$}},}\ }\href {\doibase
  10.1103/PhysRevLett.78.338} {\bibfield  {journal} {\bibinfo  {journal}
  {Physical Review Letters}\ }\textbf {\bibinfo {volume} {78}},\ \bibinfo
  {pages} {338--341} (\bibinfo {year} {1997})}\BibitemShut {NoStop}%
\bibitem [{\citenamefont {Moodenbaugh}\ \emph {et~al.}(1998)\citenamefont
  {Moodenbaugh}, \citenamefont {Wu}, \citenamefont {Zhu}, \citenamefont
  {Lewis},\ and\ \citenamefont {Cox}}]{Moodenbaugh1998}%
  \BibitemOpen
  \bibfield  {author} {\bibinfo {author} {\bibfnamefont {A.~R.}\ \bibnamefont
  {Moodenbaugh}}, \bibinfo {author} {\bibfnamefont {Lijun}\ \bibnamefont {Wu}},
  \bibinfo {author} {\bibfnamefont {Yimei}\ \bibnamefont {Zhu}}, \bibinfo
  {author} {\bibfnamefont {L.~H.}\ \bibnamefont {Lewis}}, \ and\ \bibinfo
  {author} {\bibfnamefont {D.~E.}\ \bibnamefont {Cox}},\ }\bibfield  {title}
  {\enquote {\bibinfo {title} {{High-resolution X-ray diffraction study of
  La$_{1.88--y}$Sr$_{0.12}$Nd$_y$CuO$_4$}},}\ }\href {\doibase
  10.1103/physrevb.58.9549} {\bibfield  {journal} {\bibinfo  {journal}
  {Physical Review B}\ }\textbf {\bibinfo {volume} {58}},\ \bibinfo {pages}
  {9549--9555} (\bibinfo {year} {1998})}\BibitemShut {NoStop}%
\bibitem [{\citenamefont {Berg}\ \emph {et~al.}(2007)\citenamefont {Berg},
  \citenamefont {Fradkin}, \citenamefont {Kim}, \citenamefont {Kivelson},
  \citenamefont {Oganesyan}, \citenamefont {Tranquada},\ and\ \citenamefont
  {Zhang}}]{Berg2007}%
  \BibitemOpen
  \bibfield  {author} {\bibinfo {author} {\bibfnamefont {E.}~\bibnamefont
  {Berg}}, \bibinfo {author} {\bibfnamefont {E.}~\bibnamefont {Fradkin}},
  \bibinfo {author} {\bibfnamefont {E.~A.}\ \bibnamefont {Kim}}, \bibinfo
  {author} {\bibfnamefont {S.~A.}\ \bibnamefont {Kivelson}}, \bibinfo {author}
  {\bibfnamefont {V.}~\bibnamefont {Oganesyan}}, \bibinfo {author}
  {\bibfnamefont {J.~M.}\ \bibnamefont {Tranquada}}, \ and\ \bibinfo {author}
  {\bibfnamefont {S.~C.}\ \bibnamefont {Zhang}},\ }\bibfield  {title} {\enquote
  {\bibinfo {title} {{Dynamical layer decoupling in a stripe-ordered
  high-T$_\mathrm{c}$ superconductor}},}\ }\href {\doibase
  10.1103/PhysRevLett.99.127003} {\bibfield  {journal} {\bibinfo  {journal}
  {Physical Review Letters}\ }\textbf {\bibinfo {volume} {99}},\ \bibinfo
  {pages} {1--4} (\bibinfo {year} {2007})}\BibitemShut {NoStop}%
\bibitem [{\citenamefont {Suzuki}\ and\ \citenamefont
  {Fujita}(1989)}]{Suzuki1989}%
  \BibitemOpen
  \bibfield  {author} {\bibinfo {author} {\bibfnamefont {Takashi}\ \bibnamefont
  {Suzuki}}\ and\ \bibinfo {author} {\bibfnamefont {Toshizo}\ \bibnamefont
  {Fujita}},\ }\bibfield  {title} {\enquote {\bibinfo {title} {{Structural
  phase transition in (La$_{1-x}$Ba$_x$)$_2$CuO$_{4--\delta}$}},}\ }\href
  {\doibase 10.1016/0921-4534(89)90111-1} {\bibfield  {journal} {\bibinfo
  {journal} {Physica C: Superconductivity}\ }\textbf {\bibinfo {volume}
  {159}},\ \bibinfo {pages} {111--116} (\bibinfo {year} {1989})}\BibitemShut
  {NoStop}%
\bibitem [{\citenamefont {Cox}\ \emph {et~al.}(1989)\citenamefont {Cox},
  \citenamefont {Zolliker}, \citenamefont {Axe}, \citenamefont {Moudden},
  \citenamefont {Moodenbaugh},\ and\ \citenamefont {Xu}}]{Cox1989}%
  \BibitemOpen
  \bibfield  {author} {\bibinfo {author} {\bibfnamefont {D.~E.}\ \bibnamefont
  {Cox}}, \bibinfo {author} {\bibfnamefont {P.}~\bibnamefont {Zolliker}},
  \bibinfo {author} {\bibfnamefont {J.~D.}\ \bibnamefont {Axe}}, \bibinfo
  {author} {\bibfnamefont {A.~H.}\ \bibnamefont {Moudden}}, \bibinfo {author}
  {\bibfnamefont {A.~R.}\ \bibnamefont {Moodenbaugh}}, \ and\ \bibinfo {author}
  {\bibfnamefont {Y.}~\bibnamefont {Xu}},\ }\bibfield  {title} {\enquote
  {\bibinfo {title} {{Structural studies of La2-xBaxCuO4 between 11-293 K}},}\
  }\href@noop {} {\bibfield  {journal} {\bibinfo  {journal} {Mat. Res. Soc.
  Symp. Proc.}\ }\textbf {\bibinfo {volume} {156}},\ \bibinfo {pages}
  {141--151} (\bibinfo {year} {1989})}\BibitemShut {NoStop}%
\bibitem [{\citenamefont {Sakita}\ \emph {et~al.}(1999)\citenamefont {Sakita},
  \citenamefont {Nakamura}, \citenamefont {Suzuki},\ and\ \citenamefont
  {Fujita}}]{Sakita1999}%
  \BibitemOpen
  \bibfield  {author} {\bibinfo {author} {\bibfnamefont {Shigenobu}\
  \bibnamefont {Sakita}}, \bibinfo {author} {\bibfnamefont {Fumihiko}\
  \bibnamefont {Nakamura}}, \bibinfo {author} {\bibfnamefont {Takashi}\
  \bibnamefont {Suzuki}}, \ and\ \bibinfo {author} {\bibfnamefont {Toshizo}\
  \bibnamefont {Fujita}},\ }\bibfield  {title} {\enquote {\bibinfo {title}
  {{Structural transitions and localization in La$_{2-x-y}$Nd$_y$Sr$_x$CuO$_4$
  with p$\sim\sfrac{1}{8}$}},}\ }\href {\doibase 10.1143/JPSJ.68.2755}
  {\bibfield  {journal} {\bibinfo  {journal} {Journal of the Physical Society
  of Japan}\ }\textbf {\bibinfo {volume} {68}},\ \bibinfo {pages} {2755--2761}
  (\bibinfo {year} {1999})}\BibitemShut {NoStop}%
\bibitem [{\citenamefont {Billinge}\ \emph {et~al.}(1994)\citenamefont
  {Billinge}, \citenamefont {Kwei},\ and\ \citenamefont
  {Takagi}}]{Billinge1994}%
  \BibitemOpen
  \bibfield  {author} {\bibinfo {author} {\bibfnamefont {Simon~J.L.}\
  \bibnamefont {Billinge}}, \bibinfo {author} {\bibfnamefont {George~H.}\
  \bibnamefont {Kwei}}, \ and\ \bibinfo {author} {\bibfnamefont
  {H.}~\bibnamefont {Takagi}},\ }\bibfield  {title} {\enquote {\bibinfo {title}
  {{Structural ground-state of La2CuO4 in the {LTO} phase: Evidence of local
  disorder}},}\ }\href {\doibase 10.1016/0921-4534(94)91865-1} {\bibfield
  {journal} {\bibinfo  {journal} {Physica C: Superconductivity and its
  applications}\ }\textbf {\bibinfo {volume} {235-240}},\ \bibinfo {pages}
  {1281--1282} (\bibinfo {year} {1994})}\BibitemShut {NoStop}%
\bibitem [{\citenamefont {Reehuis}\ \emph {et~al.}(2006)\citenamefont
  {Reehuis}, \citenamefont {Ulrich}, \citenamefont {Proke{\aa}}, \citenamefont
  {Gozar}, \citenamefont {Blumberg}, \citenamefont {Komiya}, \citenamefont
  {Ando}, \citenamefont {Pattison},\ and\ \citenamefont
  {Keimer}}]{Reehuis2006}%
  \BibitemOpen
  \bibfield  {author} {\bibinfo {author} {\bibfnamefont {M.}~\bibnamefont
  {Reehuis}}, \bibinfo {author} {\bibfnamefont {C.}~\bibnamefont {Ulrich}},
  \bibinfo {author} {\bibfnamefont {K.}~\bibnamefont {Proke{\aa}}}, \bibinfo
  {author} {\bibfnamefont {A.}~\bibnamefont {Gozar}}, \bibinfo {author}
  {\bibfnamefont {G.}~\bibnamefont {Blumberg}}, \bibinfo {author}
  {\bibfnamefont {Seiki}\ \bibnamefont {Komiya}}, \bibinfo {author}
  {\bibfnamefont {Yoichi}\ \bibnamefont {Ando}}, \bibinfo {author}
  {\bibfnamefont {P.}~\bibnamefont {Pattison}}, \ and\ \bibinfo {author}
  {\bibfnamefont {B.}~\bibnamefont {Keimer}},\ }\bibfield  {title} {\enquote
  {\bibinfo {title} {{Crystal structure and high-field magnetism of
  La$_2$CuO$_4$}},}\ }\href {\doibase 10.1103/PhysRevB.73.144513} {\bibfield
  {journal} {\bibinfo  {journal} {Physical Review B - Condensed Matter and
  Materials Physics}\ }\textbf {\bibinfo {volume} {73}},\ \bibinfo {pages}
  {1--8} (\bibinfo {year} {2006})}\BibitemShut {NoStop}%
\bibitem [{\citenamefont {Sapkota}\ \emph {et~al.}(2021)\citenamefont
  {Sapkota}, \citenamefont {Sterling}, \citenamefont {Lozano}, \citenamefont
  {Li}, \citenamefont {Cao}, \citenamefont {Garlea}, \citenamefont {Reznik},
  \citenamefont {Li}, \citenamefont {Zaliznyak}, \citenamefont {Gu},\ and\
  \citenamefont {Tranquada}}]{Sapkota2021}%
  \BibitemOpen
  \bibfield  {author} {\bibinfo {author} {\bibfnamefont {A.}~\bibnamefont
  {Sapkota}}, \bibinfo {author} {\bibfnamefont {T.~C.}\ \bibnamefont
  {Sterling}}, \bibinfo {author} {\bibfnamefont {P.~M.}\ \bibnamefont
  {Lozano}}, \bibinfo {author} {\bibfnamefont {Yangmu}\ \bibnamefont {Li}},
  \bibinfo {author} {\bibfnamefont {Huibo}\ \bibnamefont {Cao}}, \bibinfo
  {author} {\bibfnamefont {V.~O.}\ \bibnamefont {Garlea}}, \bibinfo {author}
  {\bibfnamefont {D.}~\bibnamefont {Reznik}}, \bibinfo {author} {\bibfnamefont
  {Qiang}\ \bibnamefont {Li}}, \bibinfo {author} {\bibfnamefont {I.~A.}\
  \bibnamefont {Zaliznyak}}, \bibinfo {author} {\bibfnamefont {G.~D.}\
  \bibnamefont {Gu}}, \ and\ \bibinfo {author} {\bibfnamefont {J.~M.}\
  \bibnamefont {Tranquada}},\ }\bibfield  {title} {\enquote {\bibinfo {title}
  {{Reinvestigation of crystal symmetry and fluctuations in La$_2$CuO$_4$}},}\
  }\href {\doibase 10.1103/PhysRevB.104.014304} {\bibfield  {journal} {\bibinfo
   {journal} {Physical Review B}\ }\textbf {\bibinfo {volume} {104}},\ \bibinfo
  {pages} {1--11} (\bibinfo {year} {2021})}\BibitemShut {NoStop}%
\bibitem [{\citenamefont {Shannon}(1976)}]{Shannon1976a}%
  \BibitemOpen
  \bibfield  {author} {\bibinfo {author} {\bibfnamefont {R.~D.}\ \bibnamefont
  {Shannon}},\ }\bibfield  {title} {\enquote {\bibinfo {title} {{Revised
  effective ionic radii and systematic studies of interatomic distances in
  halides and chalcogenides}},}\ }\href {\doibase 10.1107/S0567739476001551}
  {\bibfield  {journal} {\bibinfo  {journal} {Acta Crystallographica Section
  A}\ }\textbf {\bibinfo {volume} {32}},\ \bibinfo {pages} {751--767} (\bibinfo
  {year} {1976})}\BibitemShut {NoStop}%
\bibitem [{\citenamefont {Coelho}(2018)}]{Coelho2018}%
  \BibitemOpen
  \bibfield  {author} {\bibinfo {author} {\bibfnamefont {Alan~A}\ \bibnamefont
  {Coelho}},\ }\bibfield  {title} {\enquote {\bibinfo {title} {{TOPAS and
  TOPAS-Academic: an optimization program integrating computer algebra and
  crystallographic objects written in C++}},}\ }\href {\doibase
  https://doi.org/10.1107/S1600576718000183} {\bibfield  {journal} {\bibinfo
  {journal} {Journal of Applied Crystallography}\ }\textbf {\bibinfo {volume}
  {51}},\ \bibinfo {pages} {210--218} (\bibinfo {year} {2018})}\BibitemShut
  {NoStop}%
\bibitem [{\citenamefont {Campbell}\ \emph {et~al.}(2007)\citenamefont
  {Campbell}, \citenamefont {Evans}, \citenamefont {Perselli},\ and\
  \citenamefont {Stokes}}]{Campbell}%
  \BibitemOpen
  \bibfield  {author} {\bibinfo {author} {\bibfnamefont {Branton~J}\
  \bibnamefont {Campbell}}, \bibinfo {author} {\bibfnamefont {John S~O}\
  \bibnamefont {Evans}}, \bibinfo {author} {\bibfnamefont {Francesca}\
  \bibnamefont {Perselli}}, \ and\ \bibinfo {author} {\bibfnamefont {Harold~T}\
  \bibnamefont {Stokes}},\ }\bibfield  {title} {\enquote {\bibinfo {title}
  {{Rietveld refinement of structural distortion-mode amplitudes}},}\
  }\href@noop {} {\bibfield  {journal} {\bibinfo  {journal} {IUCr Computing
  Commission Newsletter}\ }\textbf {\bibinfo {volume} {8}},\ \bibinfo {pages}
  {81--95} (\bibinfo {year} {2007})}\BibitemShut {NoStop}%
\bibitem [{\citenamefont {Campbell}\ \emph {et~al.}(2006)\citenamefont
  {Campbell}, \citenamefont {Stokes}, \citenamefont {Tanner},\ and\
  \citenamefont {Hatch}}]{Campbell2006}%
  \BibitemOpen
  \bibfield  {author} {\bibinfo {author} {\bibfnamefont {Branton~J.}\
  \bibnamefont {Campbell}}, \bibinfo {author} {\bibfnamefont {Harold~T.}\
  \bibnamefont {Stokes}}, \bibinfo {author} {\bibfnamefont {David~E.}\
  \bibnamefont {Tanner}}, \ and\ \bibinfo {author} {\bibfnamefont {Dorian~M.}\
  \bibnamefont {Hatch}},\ }\bibfield  {title} {\enquote {\bibinfo {title}
  {{ISODISPLACE}: a web-based tool for exploring structural distortions},}\
  }\href {\doibase 10.1107/s0021889806014075607} {\bibfield  {journal}
  {\bibinfo  {journal} {Journal of Applied Crystallography}\ }\textbf {\bibinfo
  {volume} {39}},\ \bibinfo {pages} {607--614} (\bibinfo {year}
  {2006})}\BibitemShut {NoStop}%
\bibitem [{\citenamefont {Stephens}(1999)}]{Stephens1999a}%
  \BibitemOpen
  \bibfield  {author} {\bibinfo {author} {\bibfnamefont {Peter~W.}\
  \bibnamefont {Stephens}},\ }\bibfield  {title} {\enquote {\bibinfo {title}
  {{Phenomenological model of anisotropic peak broadening in powder
  diffraction}},}\ }\href {\doibase 10.1107/s0021889898006001} {\bibfield
  {journal} {\bibinfo  {journal} {Journal of Applied Crystallography}\ }\textbf
  {\bibinfo {volume} {32}},\ \bibinfo {pages} {281--289} (\bibinfo {year}
  {1999})}\BibitemShut {NoStop}%
\bibitem [{SI()}]{SI}%
  \BibitemOpen
  \href@noop {} {\emph {\bibinfo {title} {See Supplemental Material at [URL
  will be inserted by publisher] for further details of the Rietveld
  refinements and DFT calculations}}}\BibitemShut {NoStop}%
\bibitem [{\citenamefont {Jorgensen}\ \emph {et~al.}(1987)\citenamefont
  {Jorgensen}, \citenamefont {Sch{\"{u}}ttler}, \citenamefont {Hinks},
  \citenamefont {{Capone II}}, \citenamefont {Zhang}, \citenamefont {Brodsky},\
  and\ \citenamefont {Scalapino}}]{Jorgensen1987}%
  \BibitemOpen
  \bibfield  {author} {\bibinfo {author} {\bibfnamefont {J.~D.}\ \bibnamefont
  {Jorgensen}}, \bibinfo {author} {\bibfnamefont {H.~B.}\ \bibnamefont
  {Sch{\"{u}}ttler}}, \bibinfo {author} {\bibfnamefont {D.~G.}\ \bibnamefont
  {Hinks}}, \bibinfo {author} {\bibfnamefont {D.~W.}\ \bibnamefont {{Capone
  II}}}, \bibinfo {author} {\bibfnamefont {K.}~\bibnamefont {Zhang}}, \bibinfo
  {author} {\bibfnamefont {M.~B.}\ \bibnamefont {Brodsky}}, \ and\ \bibinfo
  {author} {\bibfnamefont {D.~J.}\ \bibnamefont {Scalapino}},\ }\bibfield
  {title} {\enquote {\bibinfo {title} {{Lattice Instability and
  High-T$_\mathrm{c}$ Superconductivity in La$_{2-x}$Ba$_x$CuO$_4$}},}\ }\href
  {http://link.aps.org/doi/10.1103/PhysRevLett.58.1024} {\bibfield  {journal}
  {\bibinfo  {journal} {Physical Review Letters}\ }\textbf {\bibinfo {volume}
  {58}},\ \bibinfo {pages} {1024--1027} (\bibinfo {year} {1987})}\BibitemShut
  {NoStop}%
\bibitem [{\citenamefont {Ablitt}\ \emph {et~al.}(2017)\citenamefont {Ablitt},
  \citenamefont {Craddock}, \citenamefont {Senn}, \citenamefont {Mostofi},\
  and\ \citenamefont {Bristowe}}]{Ablitt2017}%
  \BibitemOpen
  \bibfield  {author} {\bibinfo {author} {\bibfnamefont {Chris}\ \bibnamefont
  {Ablitt}}, \bibinfo {author} {\bibfnamefont {Sarah}\ \bibnamefont
  {Craddock}}, \bibinfo {author} {\bibfnamefont {Mark~S}\ \bibnamefont {Senn}},
  \bibinfo {author} {\bibfnamefont {Arash~A}\ \bibnamefont {Mostofi}}, \ and\
  \bibinfo {author} {\bibfnamefont {Nicholas~C}\ \bibnamefont {Bristowe}},\
  }\bibfield  {title} {\enquote {\bibinfo {title} {{The Origin of Uniaxial
  Negative Thermal Expansion in Layered Perovskites}},}\ }\href {\doibase
  10.1038/s41524-017-0040-0} {\bibfield  {journal} {\bibinfo  {journal}
  {Computational Materials}\ }\textbf {\bibinfo {volume} {3}},\ \bibinfo
  {pages} {44} (\bibinfo {year} {2017})}\BibitemShut {NoStop}%
\bibitem [{\citenamefont {H{\"{u}}cker}(2012)}]{Hucker2012}%
  \BibitemOpen
  \bibfield  {author} {\bibinfo {author} {\bibfnamefont {M.}~\bibnamefont
  {H{\"{u}}cker}},\ }\bibfield  {title} {\enquote {\bibinfo {title}
  {{Structural aspects of materials with static stripe order}},}\ }\href
  {\doibase 10.1016/j.physc.2012.04.035} {\bibfield  {journal} {\bibinfo
  {journal} {Physica C: Superconductivity and its Applications}\ }\textbf
  {\bibinfo {volume} {481}},\ \bibinfo {pages} {3--14} (\bibinfo {year}
  {2012})}\BibitemShut {NoStop}%
\bibitem [{\citenamefont {Pesant}\ and\ \citenamefont
  {C\^ot\'e}(2011)}]{cote2011}%
  \BibitemOpen
  \bibfield  {author} {\bibinfo {author} {\bibfnamefont {Simon}\ \bibnamefont
  {Pesant}}\ and\ \bibinfo {author} {\bibfnamefont {Michel}\ \bibnamefont
  {C\^ot\'e}},\ }\bibfield  {title} {\enquote {\bibinfo {title} {{DFT+{\it U}
  study of magnetic order in doped La$_2$CuO$_4$ crystals}},}\ }\href {\doibase
  10.1103/PhysRevB.84.085104} {\bibfield  {journal} {\bibinfo  {journal} {Phys.
  Rev. B}\ }\textbf {\bibinfo {volume} {84}},\ \bibinfo {pages} {085104}
  (\bibinfo {year} {2011})}\BibitemShut {NoStop}%
\bibitem [{\citenamefont {Sterling}\ and\ \citenamefont
  {Reznik}(2021)}]{Sterling2021}%
  \BibitemOpen
  \bibfield  {author} {\bibinfo {author} {\bibfnamefont {Tyler~C.}\
  \bibnamefont {Sterling}}\ and\ \bibinfo {author} {\bibfnamefont {Dmitry}\
  \bibnamefont {Reznik}},\ }\bibfield  {title} {\enquote {\bibinfo {title}
  {{Effect of the electronic charge gap on LO bond-stretching phonons undoped
  La$_2$CuO$_4$ calculated using LDA+{\it U}}},}\ }\href {\doibase
  10.1103/PhysRevB.104.134311} {\bibfield  {journal} {\bibinfo  {journal}
  {Phys. Rev. B}\ }\textbf {\bibinfo {volume} {104}},\ \bibinfo {pages}
  {134311} (\bibinfo {year} {2021})}\BibitemShut {NoStop}%
\bibitem [{\citenamefont {Furness}\ \emph {et~al.}(2018)\citenamefont
  {Furness}, \citenamefont {Zhang}, \citenamefont {Lane}, \citenamefont {Buda},
  \citenamefont {Barbiellini}, \citenamefont {Markiewicz}, \citenamefont
  {Bansil},\ and\ \citenamefont {Sun}}]{Furness2018}%
  \BibitemOpen
  \bibfield  {author} {\bibinfo {author} {\bibfnamefont {James~W.}\
  \bibnamefont {Furness}}, \bibinfo {author} {\bibfnamefont {Yubo}\
  \bibnamefont {Zhang}}, \bibinfo {author} {\bibfnamefont {Christopher}\
  \bibnamefont {Lane}}, \bibinfo {author} {\bibfnamefont {Ioana~Gianina}\
  \bibnamefont {Buda}}, \bibinfo {author} {\bibfnamefont {Bernardo}\
  \bibnamefont {Barbiellini}}, \bibinfo {author} {\bibfnamefont {Robert~S.}\
  \bibnamefont {Markiewicz}}, \bibinfo {author} {\bibfnamefont {Arun}\
  \bibnamefont {Bansil}}, \ and\ \bibinfo {author} {\bibfnamefont {Jianwei}\
  \bibnamefont {Sun}},\ }\bibfield  {title} {\enquote {\bibinfo {title} {{An
  accurate first-principles treatment of doping-dependent electronic structure
  of high-temperature cuprate superconductors}},}\ }\href {\doibase
  10.1038/s42005-018-0009-4} {\bibfield  {journal} {\bibinfo  {journal}
  {Communications Physics}\ }\textbf {\bibinfo {volume} {1}},\ \bibinfo {pages}
  {1--6} (\bibinfo {year} {2018})}\BibitemShut {NoStop}%
\end{thebibliography}%
\end{document}